\documentclass[twocolumn,pra,twocolumn,superscriptaddress]{revtex4}
\usepackage{color}
\usepackage{amsmath}
\usepackage{amssymb}
\usepackage{graphicx}
\usepackage{tikz}

\usepackage{float}
\usepackage{bbm}

\usepackage{epsfig} 
\usepackage{verbatim}
\usepackage{array}
\usepackage{setspace}

\usepackage[unicode=true,
 bookmarks=true,bookmarksnumbered=false,bookmarksopen=false,
 breaklinks=false,pdfborder={0 0 1},backref=false,colorlinks=true]
 {hyperref}
\hypersetup{
 linkcolor=magenta, urlcolor=blue, citecolor=blue, pdfstartview={FitH}, hyperfootnotes=false, unicode=true}
 
\usepackage{color}

\makeatletter
\@ifundefined{textcolor}{}
{%
 \definecolor{BLACK}{gray}{0}
 \definecolor{WHITE}{gray}{1}
 \definecolor{RED}{rgb}{1,0,0}
 \definecolor{GREEN}{rgb}{0,1,0}
 \definecolor{BLUE}{rgb}{0,0,1}
 \definecolor{CYAN}{cmyk}{1,0,0,0}
 \definecolor{MAGENTA}{cmyk}{0,1,0,0}
 \definecolor{YELLOW}{cmyk}{0,0,1,0}
}

\usepackage{amsfonts}\usepackage{tabularx}\usepackage{dcolumn}\usepackage{bm}\usepackage{graphicx}\usepackage{epstopdf}

\hypersetup{urlcolor=blue}

\makeatother

\newcolumntype{C}[1]{>{\centering\arraybackslash$}p{#1}<{$}}

\begin{document}

\widetext

\title{Suppression of leakage for a charge quadrupole qubit in triangular geometry}

\author{Guo Xuan Chan}
\affiliation{Department of Physics, City University of Hong Kong, Tat Chee Avenue, Kowloon, Hong Kong SAR, China, and City University of Hong Kong Shenzhen Research Institute, Shenzhen, Guangdong 518057, China}
\author{Xin Wang}
\email{x.wang@cityu.edu.hk}
\affiliation{Department of Physics, City University of Hong Kong, Tat Chee Avenue, Kowloon, Hong Kong SAR, China, and City University of Hong Kong Shenzhen Research Institute, Shenzhen, Guangdong 518057, China}
\date{\today}

\begin{abstract}
We present a simple yet effective strategy to suppress the leakage in the charge quadrupole qubit system by having the dots in a triangle as opposed to the linear geometry originally conceived. We find that the tunnel coupling between the two outmost dots are amplified in triangular triple dots, which consequently reduces leakage by separating the leaked state and qubit states. We have found that the leakage can be suppressed by as much as five orders of magnitude when the dots form an equilateral triangle, with further improvement of control fidelities possible if composite pulses are applied.
\end{abstract}
\maketitle

\textit{Introduction.} Semiconductor quantum dots are promising candidates to realize quantum computing. In the simplest proposal of a double-quantum-dot charge qubit, an electron can occupy either one dot or the other, serving as the two logical states \cite{Hayashi.03}. Universal single-qubit operations of the charge qubit can be achieved by alternating the relative energy between the two dots \cite{Fujisawa.04,Fujisawa.06}, and two-qubit gates can be performed using the electrostatic coupling between them \cite{Shinkai.09,Li.15,Ward.16}. While this charge qubit can have very fast gate operations, it at the same time suffers heavily from charge noises, leading to very short coherence times \cite{Gorman.05,Petersson.10,Dovzhenko.11,Shi.13,Penfold.17,Wang.17CQ}. Various proposals have been put forward to improve the control of charge qubits for quantum information processing, including ultrafast gates \cite{Cao.13}, ac control around certain sweet spots \cite{Kim.15}, or hybrid qubits that combine advantages of manipulating charge and spin states \cite{Cao.16,Wong.16,Serina.17,Yang.17Wisc,Wang.17HYB}.

Very recently, a new type of charge qubit, called the charge quadrupole (CQ), has been proposed \cite{Friesen.17,Ghosh.17,Kornich.18}. In the CQ qubit, an electron resides in a  triple-quantum-dot system, giving rise to three electron states.
In the lateral geometry, the tunneling between the two outermost dots are neglected, and one is left with only two barriers to address \cite{Friesen.17,Yang.18}.  It is argued that if the energies of the two barriers can be kept equal for the duration of the operation, and the same can be done for the the energies of the two outmost dots, we are left with a decoherence free subspace against charge noise with which the fidelity may be substantially improved \cite{Friesen.17}. However, practically such a stringent requirement is difficult to satisfy, and even a slight deviation would cause considerable leakage as the leaked state is energetically very close to the qubit states. In order to suppress the leakage, composite pulses are proposed \cite{Ghosh.17,Sun.17}.

In this Letter, we present a simple yet effective strategy to suppress the leakage thus protecting the decoherence free subspace. We propose that the triple dots can alternatively be fabricated in a triangle, in which the tunneling between the two outmost dots are amplified. This tunnel coupling effectively separates the leaked state and qubit states,  reducing leakage. We performed analytical and numerical analyses of the effect of the introduced tunnel coupling on the system, and found that the leakage can be suppressed by as much as five orders of magnitude. Further improvement of control fidelities is possible when composite pulses are applied to triangular triple quantum dots.

\begin{figure} [b]
	\centering
	\includegraphics[width=0.75\columnwidth]{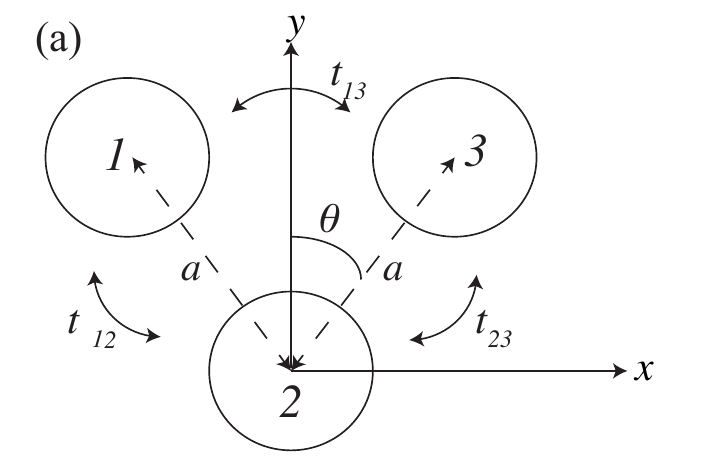}
	\centering	
	\includegraphics[width=0.7\columnwidth]{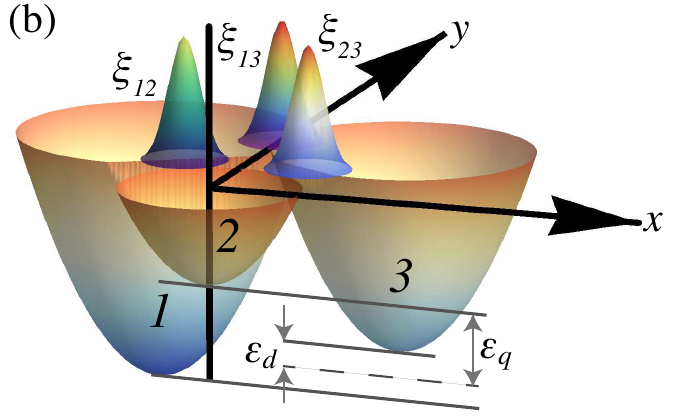}
	\caption{(a) A schematic graph of the triangular triple dots viewing from $+z$ direction. (b) The confinement potential. $\xi_{jk}$ is the barrier height between the $j^{\rm th}$ and $k^{\rm th}$ dot that is used in the Hund-Mulliken calculation. $\varepsilon_q$ and $\varepsilon_d$ are the quadrupolar and dipolar detuning energies respectively.}
	\label{fig:triangle-dot-figure}
\end{figure}

\textit{Model.} A triple-quantum-dot system in triangular geometry is shown in Fig.~\ref{fig:triangle-dot-figure}, where the dots are labelled by 1, 2, 3, and the bending angle $\theta$ controls the shape of the triangle. When $\theta=90^\circ$, the system is identical to the lateral triple dots considered in \cite{Friesen.17}. When only one electron is present in the system, we may define a set of ``position basis'' as $\{|100\rangle = c^\dagger_1 |\mathrm{vac}\rangle,|010\rangle = c^\dagger_2 |\mathrm{vac}\rangle,|001\rangle = c^\dagger_3 |\mathrm{vac}\rangle\}$, where $c^\dagger_j$ creates an electron on the $j^\mathrm{th}$ dot and $|\mathrm{vac}\rangle$ is the vacuum state. Considering the fact that the electron can hop between any two dots, the Hamiltonian can be written as 
\begin{equation} 
	\widetilde{H}_{CQ}=
	\left(
	\begin{array}{ccc}
		\varepsilon_1 & -t_{12} & -t_{13} \\
		-t_{12} & \varepsilon_2 & -t_{23} \\
		-t_{13} & -t_{23} & \varepsilon_3 \\
	\end{array}
	\right),
	\label{eq:ham-posBases}
\end{equation}
where $\varepsilon_j$ is the electron energy in the $j^\mathrm{th}$ dot and $t_{jk}$ is the tunneling energy between the $j^\mathrm{th}$ and $k^\mathrm{th}$ dot. Comparing to the lateral-triple-dot case considered in \cite{Friesen.17}, the additional terms involving $t_{13}$ have significant effects on reduction of leakage as shall be demonstrated below.

\begin{figure} [t]
	\includegraphics[width=0.95\columnwidth]{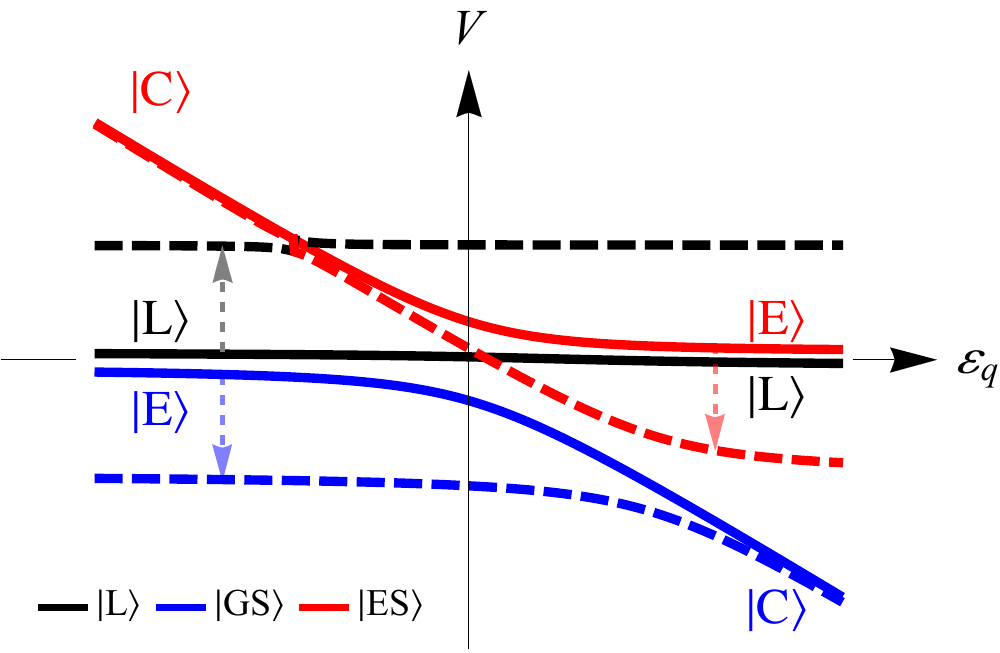}
	\caption{Schematic figure of the energy spectrum of $H_{\rm CQ}$  [Eq.~\eqref{eq:ham-CEL}] as a function of $\varepsilon_q$. The blue (red, black) solid line shows the eigenenergy of state $|GS\rangle$ ($|ES\rangle$, $|\widetilde{L}\rangle$) when $|t_{13}| = 0$. The corresponding dashed lines show the levels when $t_{13} > 0$. The arrows indicate how the energies of the states are modified by $t_{13}$.}
	\label{fig:spectra}
\end{figure}

The charge quadrupole (CQ) qubit is defined in the ``even-odd'' bases, 
$\{|C\rangle = |010\rangle,|E\rangle = (|100\rangle+|001\rangle)/\sqrt{2},|L\rangle = (|100\rangle-|001\rangle)/\sqrt{2}\}$, where $|C\rangle$ and $|E\rangle$ are the logical bases, and $|L\rangle$ a leaked state. Under this set of bases, the Hamiltonian of a CQ qubit is
\begin{equation}
H_{\rm CQ}= 	
	\begin{pmatrix}
	-\varepsilon_q & -t & 0 \\
	-t & -t_{13} & 0 \\
	0 & 0 & t_{13} \\
	\end{pmatrix}
	+H_{\rm leak},
\label{eq:ham-CEL}
\end{equation}
where
\begin{equation}
H_{\rm leak}= 	
	\begin{pmatrix}
	0 & 0 & \Delta t \\
	0 & 0 & \varepsilon_d \\
	\Delta t & \varepsilon_d & 0 \\
	\end{pmatrix},
\end{equation}
$t = (t_{12} + t_{23})/\sqrt{2}$, $\Delta t = (t_{23}-t_{12})/\sqrt{2}$, $\varepsilon_d = (\varepsilon_1 - \varepsilon_3)/2 $, and $\varepsilon_q = (\varepsilon_1 + \varepsilon_3)/2-\varepsilon_2 $ (an overall energy shift of $(\varepsilon_1 + \varepsilon_3)/2$ has been removed). When $\Delta t=\varepsilon_d=0$, the leaked state has an energy of $t_{13}$ and completely decouples from the qubit states. The CQ qubit is then operated by varying $t$ and $\varepsilon_q$. Nevertheless, maintaining $\Delta t=\varepsilon_d=0$ is a difficult practice. One must keep $t_{12} = t_{23}$ and $\varepsilon_1 = \varepsilon_3$ for the duration of an operation, requiring extremely high precision in control, and a slight deviation from the equalities would cause leakage.

For the convenience of later discussions, we define leakage as follows. We denote the eigenstates of the noisy Hamiltonian [Eq.~\eqref{eq:ham-CEL}]  as $|GS\rangle$,$|ES\rangle$ and $|\widetilde{L}\rangle$. $|GS\rangle$ and  $|ES\rangle$ are the ground state and excited state respectively that are used to encode qubit states, while $|\widetilde{L}\rangle$ is the leaked state with mostly $|L\rangle$ character. The former two states can be expressed as
\begin{equation}
\begin{split}
|GS\rangle &= \alpha_{GS} |C\rangle + \beta_{GS} |E \rangle + \gamma_{GS} |L\rangle, \\
|ES\rangle &= \alpha_{ES} |C\rangle + \beta_{ES} |E \rangle + \gamma_{ES} |L\rangle.
\end{split}
\label{eq:eigenstates}
\end{equation}
The leakage is then defined as the probability of $|L\rangle$ in $|GS\rangle$ or $|ES\rangle$, i.e. $\eta_{|GS\rangle} = |\gamma_{GS}|^2$, $\eta_{|ES\rangle} = |\gamma_{ES}|^2$.

Below, we shall demonstrate that increasing $t_{13}$ moves the leaked state far away in energy and substantially reduces leakage. 

\begin{figure}[t]
	\centering
	\includegraphics[width=1\columnwidth]{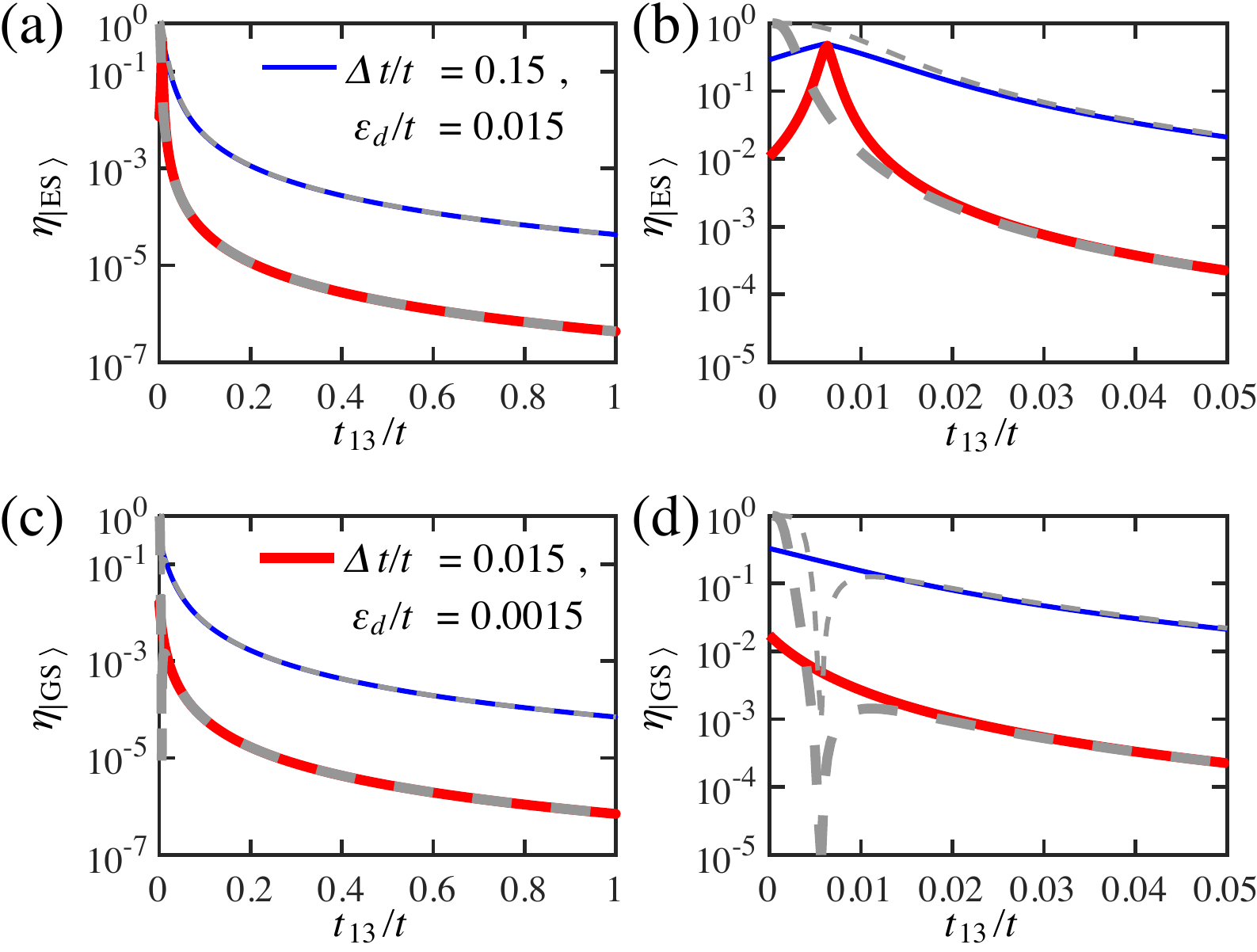}
	\caption{Leakage as a function of tunneling rate $t_{13}$ where $\Delta t >$ 0 and $\varepsilon_d>$ 0 for (a) $|ES\rangle$ at $\varepsilon_q=800 \mu $eV and (c) $|GS\rangle$ at $\varepsilon_q=-800 \mu $eV. $t=10 \mu $eV. Solid (dashed) line shows the numerical (analytical) solution. (b) ((d)) is a zoom-in version of the small $\varepsilon_q$ range of (a) ((c)).}
	\label{fig:leakage-dt-ed-nonzero}
\end{figure}

\textit{Results.}  Figure~\ref{fig:spectra} shows qualitatively the energy levels of the Hamiltonian $H_{\rm CQ}$ [Eq.~\eqref{eq:ham-CEL}]. When $|\varepsilon_q| \gg 0$ (away from the avoided crossing), the eigenstates are predominantly $|C\rangle$, $|E\rangle$ or $|L\rangle$ states, respectively. Also in this range, the energy of the state which is mainly $|E\rangle$ character becomes asymptotically close to that of $|L\rangle$, causing heavy leakage when $\Delta t$ or $\varepsilon_d$ are nonzero. To further understand how $t_{13}$ changes leakage, we have performed analytical calculations on Eq.~\eqref{eq:ham-CEL} treating $H_{\rm leak}$ as a perturbation, and details can be found in the Supplemental Material \cite{sm}. We have shown that the energy difference between $|GS\rangle$ and $|\widetilde{L}\rangle$ (denoted as $\Delta E_{GS}$), and that between $|ES\rangle$ and $|\widetilde{L}\rangle$ ($\Delta E_{ES}$) can be expressed in the limit of $|\varepsilon_q|\gg 0$ as \cite{sm}:
\begin{equation}
\begin{split}
\Delta E_{GS}=E_{GS}-E_L&\approx-2t_{13}-\delta , \quad(\varepsilon_q < 0)\\
\Delta E_{ES} = E_{ES} - E_L &\approx-2t_{13}+\delta, \quad(\varepsilon_q>0)
\end{split}
\label{eq:energydiff}
\end{equation}
where $\delta=t^2/|\varepsilon_q-t_{13}|$ is typically small when $t < t_{13} \ll |\varepsilon_q|$. Therefore, the introduction of $t_{13}$ does two things: it moves the energy of the $|E\rangle$ state down by $-t_{13}$ (away from the avoided crossing), while moving the energy of the $|L\rangle$ state up by $t_{13}$, causing an energy difference of $2t_{13}$ between them. The larger $t_{13}$ is, the further apart the two states are, and the smaller the leakage would be. This effect is shown as the dashed lines in Fig.~\ref{fig:spectra} where the arrows indicate the effect of $t_{13}$.

Figure~\ref{fig:leakage-dt-ed-nonzero} compares the numerical and perturbation theory calculations of the leakage as a function of $t_{13}/t$ when both $\Delta t$ and $\varepsilon_d$ are non-zero. The numerical solutions (shown as solid lines) are derived from a direct diagonalization of Eq.~\eqref{eq:ham-CEL}, while the perturbation theory results are from Sec. I of the Supplemental Material \cite{sm}. From Fig.~\ref{fig:leakage-dt-ed-nonzero}(a) and (c) we see that increasing $t_{13}$ indeed reduces leakage drastically. When $t_{13}$ is about 20\% of $t$, the leakage can be reduced by roughly three orders of magnitude for the case with $\Delta t/t=0.15$ and $\varepsilon_d/t=0.015$, and five orders of magnitude for the case with $\Delta t/t=0.015$ and $\varepsilon_d/t=0.0015$. When the magnitude of $t_{13}$ is approaching $t$ (which is experimentally feasible when the three dots form a equilateral triangle as shall be discussed later), the reduction of leakage is more than three orders of magnitude for the case with larger $\Delta t$ and $\varepsilon_d$, and about five orders of magnitude for the case with smaller $\Delta t$ and $\varepsilon_d$.

It is also interesting to note the existence of a cusp at $t_{13}/t\approx0.006$ in Fig.~\ref{fig:leakage-dt-ed-nonzero}(b), a zoomed-in version of Fig.~\ref{fig:leakage-dt-ed-nonzero}(a). This happens because when $t_{13}=0$, $|ES\rangle$ is slightly above $|L\rangle$ in energy. When $t_{13}$ is increased to $t_{13}/t\approx0.006$ at these parameters, $|ES\rangle$ first moves down closer to $|L\rangle$, and then moves further away from $|L\rangle$ as $t_{13}$ is further increased. Therefore the leakage first increases, then decreases substantially. Since the increase of leakage only happens for very small $t_{13}$, we believe that it does not affect our main results. We also note that at this very point the non-degenerate perturbation theory would fail, thus the disagreement between analytical and numerical results in Fig.~\ref{fig:leakage-dt-ed-nonzero}(b) and (d) \cite{leak}. 

We note that $\Delta t$ and $\varepsilon_d$ do not contribute equally to the leakage. The leakage can be approximately shown to depend quadratically on both $\Delta t/t_{13}$ and $\varepsilon_d/t_{13}$ with different prefactors \cite{sm}. For a more detailed discussion as well as results of leakage when only one of $\Delta t$ and $\varepsilon_d$ is allowed to be nonzero, see \cite{sm}. In any case, the leakage to $|L\rangle$ is substantially reduced when $t_{13}$ is increased.

We further investigate how $t_{13}$ can be increased by simply altering the geometry of the triple dots. This information is not in the model Hamiltonian Eq.~\eqref{eq:ham-posBases}, and we must calculate using a microscopic theory. In this work, we calculate various parameters in Eq.~\eqref{eq:ham-posBases} using standard Hund-Mulliken approximation, which retains the lowest energy level in each dot. The details of the calculation can be found in \cite{sm}. While $\varepsilon_d$ can be tuned in the calculation by simply changing the dot energies, $\Delta t$ is varied, in the microscopic calculations, by altering the heights of the tunneling barriers between dots, $\xi_{12}$, $\xi_{23}$ and $\xi_{13}$ as indicated in Fig.~\ref{fig:triangle-dot-figure}(b). Nevertheless, our calculations reveal that differences between $\xi_{12}$ and $\xi_{23}$, which we denote as $\Delta \xi=(\xi_{12}-\xi_{23})/2$, will indirectly gives rise to dipolar detuning noise $\varepsilon_d$. We therefore will not separate $\Delta t$ noise but instead use $\Delta\xi$  in the relevant discussions. We also define $\xi=(\xi_{12}+\xi_{23})/2$ for convenience. 

\begin{figure} [t]
	\centering
	\includegraphics[width=0.7\columnwidth]{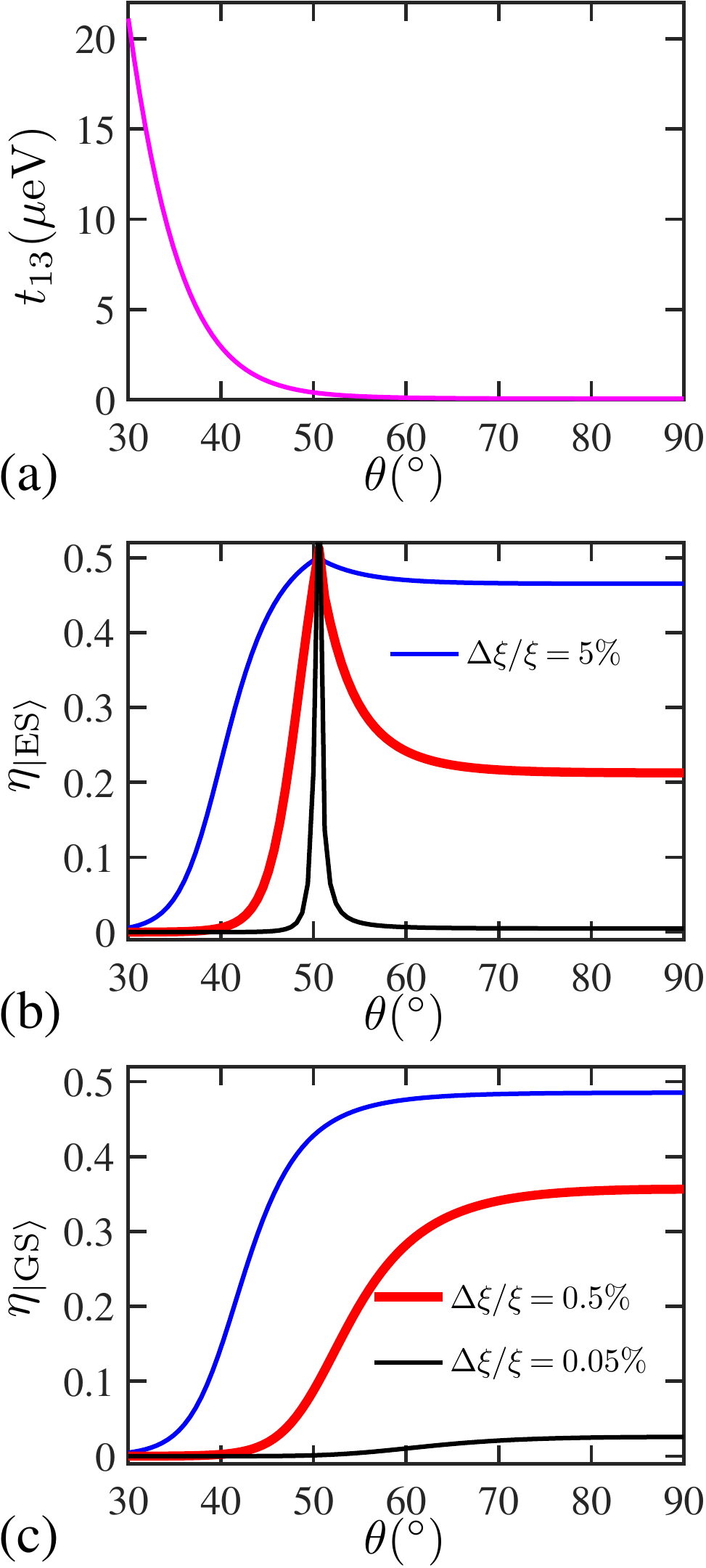}
	\caption{(a) $t_{13}$ as a function of bending angle $\theta$ for $\varepsilon_q = 800\mu$eV. (b), (c): Leakage as a function of bending angle $\theta$ due to $\Delta \xi$ for (b) $|ES\rangle$ at $\varepsilon_q=800 \mu $eV and (c) $|GS\rangle$ at $\varepsilon_q=-800 \mu $eV. Blue (red, black) line shows result for $\Delta\xi/\xi=5\%$, (0.5\%, 0.05\%), which corresponds to $\Delta t/t =21\%$, (2.1\%, 0.21\%) and $\varepsilon_d/t = 1.3$, (13\%, 0.13\%) in the language of Eq.~\eqref{eq:ham-CEL}. The parameters are $a=150$nm, $\hbar \omega_0 = 0.7$meV, $\xi = 2.2$meV, $\xi_{13}=0$, $m^*=0.067m_e$ for GaAs.}
	\label{fig:leakage-HM-dt}
\end{figure}

Figure~\ref{fig:leakage-HM-dt}(a) shows the magnitude of $t_{13}$ as a function of the bending angle $\theta$ (cf. Fig.~\ref{fig:triangle-dot-figure}). For $60^{\circ} < \theta < 90^{\circ}$, $t_{13}$ is negligibly small but then increases rapidly when $\theta < 60^{\circ}$. We also note that $\theta < 30^{\circ}$ is unphysical due to the fact that dot 1 and dot  3 start to overlap substantially and our triple-dot model is no longer valid. Therefore the results are not shown in that range. Fig.~\ref{fig:leakage-HM-dt}(b) and (c) show the leakage $\eta$ as functions of the bending angle $\theta$ for different levels of $\Delta\xi/\xi$. Since $t_{13}$ increases as $\theta$ is reduced, these figures are best viewed from right to left. Fig.~\ref{fig:leakage-HM-dt}(b) shows the leakage for the excited state. For the range of $60^{\circ} \lesssim \theta \lesssim 90^{\circ}$ the leakage remains mostly constant, which is consistent with the results shown in Fig.~\ref{fig:leakage-HM-dt}(a) that $t_{13}$ is almost zero. As $\theta$ approaches about $50^{\circ}$, the leakage increases because $t_{13}$ is increased to a small positive value which brings the excited and the leaked state very close in energy, also consistent with the observation in Fig.~\ref{fig:leakage-dt-ed-nonzero}(b). After that as $\theta$ is reduced below $50^{\circ}$, $t_{13}$ rapidly increases and the leakage is drastically suppressed. Overall, we see that for lateral triple dots ($\theta=90^{\circ}$) the leakage can be as large as 0.47 for $
\Delta\xi/\xi=5\%$, but it can be suppressed by 80\% when $\theta\approx37^{\circ}$, and further down to a very small value ($\approx 3 \times 10^{-6}$) when $\theta$ approaches $30^{\circ}$. Fig.~\ref{fig:leakage-HM-dt}(c) shows the leakage of the ground state as a function of $\theta$. The leakage monotonically decreases as $\theta$ is decreased from $90^{\circ}$ to $30^{\circ}$ because a positive $t_{13}$ brings the ground state further away from the leaked state without crossing it, as is clear from Fig.~\ref{fig:spectra}. It is again obvious that when $\theta\lesssim50^{\circ}$ the reduction of leakage is substantial.

\begin{figure} [t]
	\includegraphics[width=0.9\columnwidth]{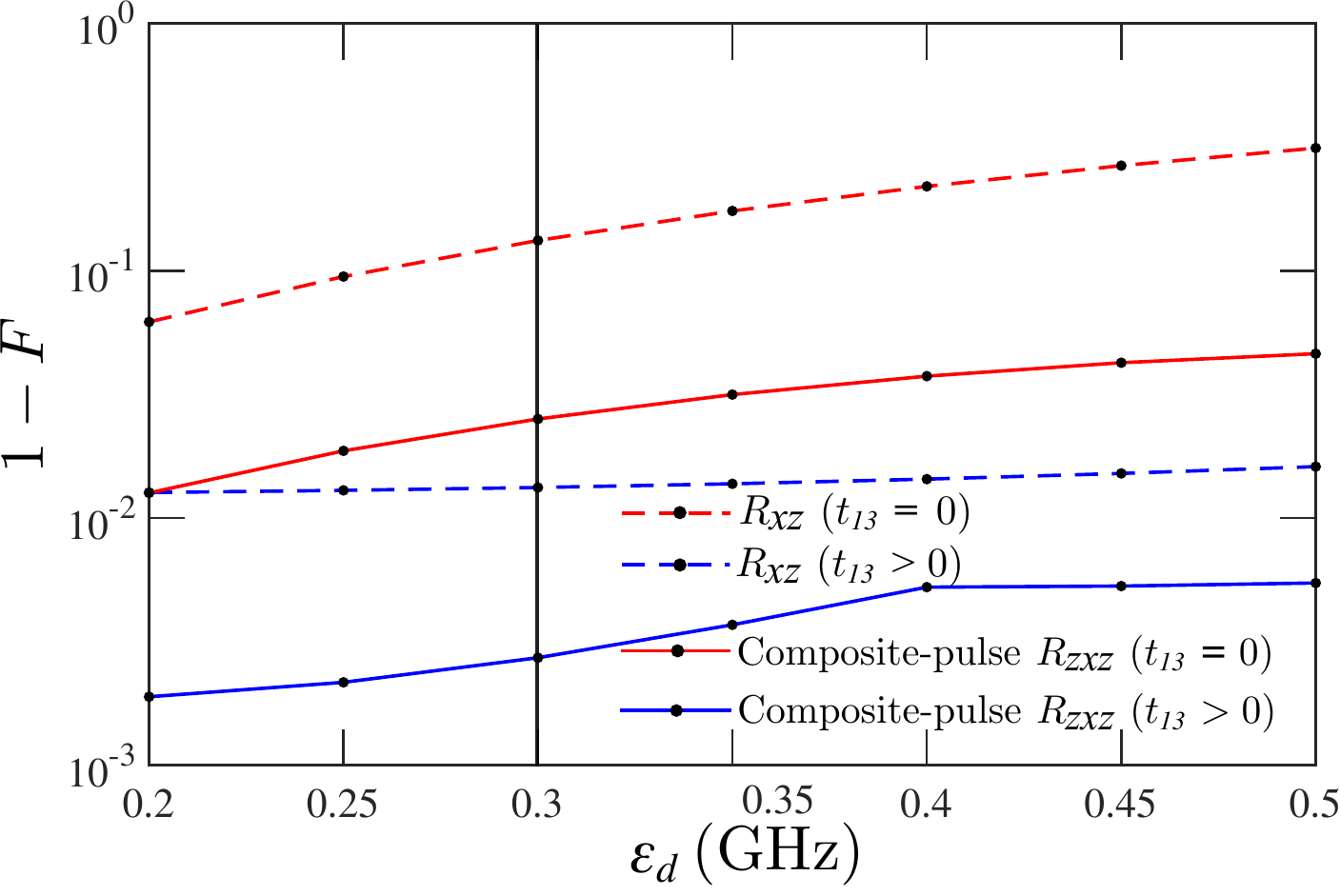}
	\caption{Infidelities of $R_{\hat{x}+\hat{y}}(\frac{\pi}{2}) $ as functions of $\varepsilon_d$ for linear (red/upper two lines, $t_{13} = 0$) and triangular triple-quantum dot (blue/lower two lines, $t_{13} > 0$). Black-filled circles are the maximally minimized infidelity for numerically optimized pulses corresponding to the given $\varepsilon_d$. The vertical black solid line indicates a typical value of $\varepsilon_d$ that is consistent with recent experiments \cite{Mi.17}. Two types of pulse sequences are compared, the naive two-step rotation, $R_{\hat{z}}(\varphi)R_{\hat{x}}(\zeta)$ (dashed lines), and the leakage suppressing composite-pulse, $R_{\hat{z}}(\varphi)R_{\hat{x}}(\zeta)R_{\hat{z}}(\varphi)$ (solid lines).}
	\label{fig:infidelityRotation}
\end{figure}

From Fig.~\ref{fig:leakage-HM-dt}(b) and (c), it is also remarkable that for $\Delta\xi/\xi=0.5\%$ the level of leakage far exceeds that of $\Delta t/t=0.15$ shown in Fig.~\ref{fig:leakage-dt-ed-nonzero}. This is due to the fact that $\Delta\xi$ not only gives rise to $\Delta t$ but also $\varepsilon_d$, which for this particular case is $\varepsilon_d/t=0.13$, a very large value. As is shown in \cite{sm}, $\varepsilon_d$ contributes more than $\Delta t$ with a much larger prefactor, therefore the leakage in this case is very large. To complete the discussion, we have also performed calculations where $\varepsilon_d$ is varied and $\Delta\xi$ is kept at 0 (note that $t_{12}$ and $t_{23}$ values are still affected), and the results are shown in \cite{sm}.

An alternative strategy to suppress leakage is the use of composite pulses with appropriately chosen parameters \cite{Ghosh.17}. We compare the method to ours and also study the benefit when we combine the two methods, i.e. implementing composite pulses with a nonzero $t_{13}$. We briefly review the key points of the composite pulse method in \cite{sm}, together with details of our implementation. We take the rotation of $\pi/2$ around axis $\hat{x}+\hat{y}$, $R_{\hat{x}+\hat{y}}(\pi/2)$, as an example.  This operation can either be achieved by a two-piece pulse-sequence, 
$R_{\text{xz}}(\varphi,\zeta) = R_{\hat{z}}(\varphi)R_{\hat{x}}(\zeta)$, if no correction to noises is desired, or a $z$-$x$-$z$ sequence $R_{\text{zxz}}(\varphi,\zeta) = R_{\hat{z}}(\varphi)R_{\hat{x}}(\zeta)R_{\hat{z}}(-\varphi)$, which can correct noises with appropriately chosen parameters \cite{Ghosh.17,sm}. We have also considered a finite rise time of 50 ps for every pulse in implementation \cite{sm}. Fig.~\ref{fig:infidelityRotation} shows the infidelity of different implementations of $R_{\hat{x}+\hat{y}}(\pi/2)$. The dashed lines show results of the naive $R_{\text{xz}}$ sequence, while solid lines the $R_{\text{zxz}}$ composite pulse sequence. Red (upper two) lines are results with $t_{13}=0$, and $t_{13} = 7.5 $ GHz ($\approx 30$ $\mu$eV) for blue (lower two) lines. We see that while the composite pulse sequences indeed reduce the leakage and consequently improves the infidelity, a nonzero $t_{13}$ is more effective: the results from the $R_{\text{xz}}$ sequence with $t_{13} = 7.5 $ GHz has lower error than results from $R_{\text{zxz}}$ sequence with $t_{13}=0$ for the range of $\varepsilon_d$ considerred. Nevertheless, the combination of the two methods, i.e. using composite pulse sequence while keeping $t_{13}>0$, yields the lowest error, which is of the order of $10^{-3}$.

\textit{Conclusions.} We suggest a simple yet effective strategy for protecting the computational subspace of a CQ qubit implemented in a triple-quantum-dot device, i.e by fabricating triple quantum dots in a triangle. The introduced $t_{13}$ as a consequence of the triangular architecture can lift the energy separation of two energetically close states ($|L\rangle$ and $|E\rangle$)  thus substantially reduces the leakage. Microscopic calculation reveals that leakage reduction becomes evident when the bending angle $\theta\lesssim 50^\circ$, and when $\theta=30^\circ$ (i.e.~the three dots forming a equilateral triangle), the leakage is suppressed by several orders of magnitude compared to those in a lateral array as originally conceived. Our results should help in achieving precise control of a CQ qubit, a promising candidate to realize quantum information processing in semiconductor quantum devices. 

\textit{Acknowledgements.} This work is supported by the Research Grants Council of the Hong Kong Special Administrative Region, China (Grant Nos.~CityU 21300116, CityU 11303617), the National Natural Science Foundation of China (Grant Nos.~11604277, 11874312), and the Guangdong Innovative and Entrepreneurial Research Team Program (Grant No.~2016ZT06D348).

%


\onecolumngrid

\vspace{.4cm}
\begin{center}
	{\bf\large Supplementary material}
\end{center}
\vspace{0.2cm}

\setcounter{secnumdepth}{3}  
\setcounter{equation}{0}
\setcounter{figure}{0}
\setcounter{table}{0}
\renewcommand{\theequation}{S-\arabic{equation}}
\renewcommand{\thefigure}{S\arabic{figure}}
\renewcommand{\thetable}{S-\Roman{table}}
\renewcommand\figurename{Supplementary Figure}
\renewcommand\tablename{Supplementary Table}

\newcolumntype{M}[1]{>{\centering\arraybackslash}m{#1}}
\newcolumntype{N}{@{}m{0pt}@{}}

\makeatletter \renewcommand\@biblabel[1]{[S#1]} \makeatother


In this Supplemental Material we provide necessary details complementary to results shown in the main text.

\section{Calculation of leakage from perturbation theory} \label{sec:analytical}

In this section we calculate the leakage from non-degenerate perturbation theory. As discussed in the main text, leakage is caused either by  $\varepsilon_d\neq0$ or $\Delta t\neq0$. When $\varepsilon_d=\Delta t=0$,  the eigenvalues  of the Hamiltonian $H_\mathrm{CQ}$ [Eq.~\eqref{eq:ham-CEL} in the main text] are
\begin{subequations}
	\begin{align}
		E_{GS} &= \frac{1}{2}\left[ -\left( \varepsilon_q+t_{13} \right)-|\varepsilon_q-t_{13}|\sqrt{1+\alpha^2}\right], \\
		E_{ES} &= \frac{1}{2}\left[ -\left( \varepsilon_q+t_{13} \right)+|\varepsilon_q-t_{13}|\sqrt{1+\alpha^2}\right], \\
		E_L &= t_{13}, 
	\end{align}
\end{subequations}
and its eigenvectors (written in $\{|C\rangle,|E\rangle,|L\rangle\}$ basis) are
\begin{subequations}
	\begin{align}
		|GS\rangle^{(0)} &=\left( \frac{\beta+\gamma}{\chi \sqrt{\alpha^2+\left(1+\beta\gamma\right)^2}},\frac{|\alpha|}{\sqrt{\alpha^2+\left(1+\beta\gamma\right)^2}},0\right), \\
		|ES\rangle^{(0)} &= \left( \frac{\beta-\gamma}{\chi \sqrt{\alpha^2+\left(1-\beta\gamma\right)^2}},\frac{|\alpha|}{\sqrt{\alpha^2+\left(1-\beta\gamma\right)^2}},0\right), \\
		|L\rangle^{(0)} &= \left(0,0,1\right),
	\end{align}
\end{subequations}
where $\alpha = |2t/(\varepsilon_q-t_{13})|$, $\beta=\mathrm{sgn}(\varepsilon_q-t_{13})$,  $\chi=\mathrm{sgn}(t)$ and $\gamma=\sqrt{1+\alpha^2}$. We have kept $\chi$ to cover the case in which $t$ may become negative, although in our work $t$ is always positive ($\chi = 1$). The superscript $(0)$ indicates that these are zeroth order results.

Either a nonzero $\varepsilon_d$ or $\Delta t$ causes leakage. We calculate their effects on the leakage using non-degenerate perturbation theory. Therefore, the analytical results shown below are valid when  $\varepsilon_d$ and $\Delta t$ are small, while $|t_{13}|$ is reasonably large such that the eigenstates are non-degenerate.

Perturbation theory suggests that the leakage is inversely proportional to the distance in energy between the two states concerned, i.e. 
$\eta_{|GS\rangle} \propto 1/(E_{GS}-E_L)$, and $\eta_{|ES\rangle} \propto 1/(E_{ES}-E_L)$. We have
\begin{align}
	\Delta E_{GS} &= E_{GS} - E_L \notag\\
	&= \frac{1}{2} \left\{-\left(\varepsilon_q+t_{13}\right)-\left[-\left(\varepsilon_q-t_{13}\right)\sqrt{1+\alpha^2}\right] \right\}-t_{13}  \qquad \qquad (\varepsilon_q < 0)\notag\\
	&\approx \frac { 1 }{ 2 } \left[ -\varepsilon _{ q }-t_{ 13 }+\left( \varepsilon _{ q }-t_{ 13 } \right) \left( 1+\frac{\alpha^2}{2}\right) \right] -t_{13}\notag
	\\
	&=-2t_{13}-\delta,
\end{align}
and
\begin{align}
	\Delta E_{ES} &= E_{ES} - E_L \notag\\
	&=\frac{1}{2} \left\{-\left(\varepsilon_q+t_{13}\right)+\left[\left(\varepsilon_q-t_{13}\right)\sqrt{1+\alpha^2}\right] \right\}-t_{13} \qquad \qquad (\varepsilon_q > 0)\notag\\
	&\approx \frac { 1 }{ 2 } \left[ -\varepsilon _{ q }-t_{ 13 }+\left( \varepsilon _{ q }-t_{ 13 } \right) \left( 1+\frac{\alpha^2}{2}\right) \right]-t_{13}\notag \\
	&=-2t_{13}+\delta,
	\label{eq:energydifference}
\end{align}
where $\delta= t^2/|\varepsilon_q-t_{13}|$. These are Eq.~\eqref{eq:energydiff} in the main text. The approximation is made under the assumption that $|\varepsilon_q| \gg t_{13}$ and $|\varepsilon_q| \gg t$, which is valid as the leakage becomes more prominent in large $|\varepsilon_q|$ regime.

To the first order perturbation correction, the eigenstates are (up to $\mathcal{O}[\alpha]^3$):
\begin{subequations}
	\begin{align}
		\begin{split}
			|GS\rangle^{(1)}&= |GS\rangle^{(0)}+\Delta t|GS\rangle^{(1)}_{\Delta t}+\varepsilon_d|GS\rangle^{(1)}_{\varepsilon_d}\\
			&=\frac{1}{c}\left( \frac { 1 }{ 2\chi  } \alpha -\frac { 3 }{ 16\chi  } \alpha ^{ 3 },
			1-\frac { 1 }{ 8 } \alpha ^{ 2 },
			-\left\{ \left[ \frac { \chi  }{ 4t_{ 13 } } \alpha +\frac { \left( \varepsilon _{ q }-4t_{ 13 } \right) \chi  }{ 32t_{ 13 }^{ 2 } } \alpha ^{ 3 } \right] \Delta t+\left[ \frac { 1 }{ 2t_{ 13 } } +\frac { \varepsilon _{ q }-2t_{ 13 }}{ 16t_{ 13 }^{ 2 } }\alpha ^{ 2 }  \right] \varepsilon _{ d } \right\}  \right),
		\end{split}\\
		\begin{split}
			|ES\rangle^{(1)}&=|ES\rangle^{(0)}+\Delta t|ES\rangle^{(1)}_{\Delta t}+\varepsilon_d|ES\rangle^{(1)}_{\varepsilon_d}\\
			&=\frac{1}{c}\left( -\frac { 1 }{ 2\chi  } \alpha +\frac { 3 }{ 16\chi  } \alpha ^{ 3 },1-\frac { \alpha ^{ 2 } }{ 8 } ,\left[ \frac { \chi  }{ 4t_{ 13 } } \alpha +\frac { \left( \varepsilon _{ q }-4t_{ 13 } \right) \chi  }{ 32t_{ 13 }^{ 2 } } \alpha ^{ 3 } \right] \Delta t-\left[ \frac { 1 }{ 2t_{ 13 } } +\frac { \varepsilon _{ q }-2t_{ 13 } }{ 16t_{ 13 }^{ 2 } }  \right] \varepsilon _{ d } \right),
		\end{split}
	\end{align}
\end{subequations}
where $c$ is the normalization constant, and the subscript $\Delta t$ ($\varepsilon_d$) indicates correction arising from $\Delta t$ ($\varepsilon_d$). We note that since the leakage is only based on the coefficient of $|L\rangle$, for which the second order correction is zero, our discussion of leakage is in fact valid to the second order of $\Delta t$ and $\varepsilon_d$.

\begin{figure}[t]
	\centering
	\includegraphics[width=0.45\columnwidth]{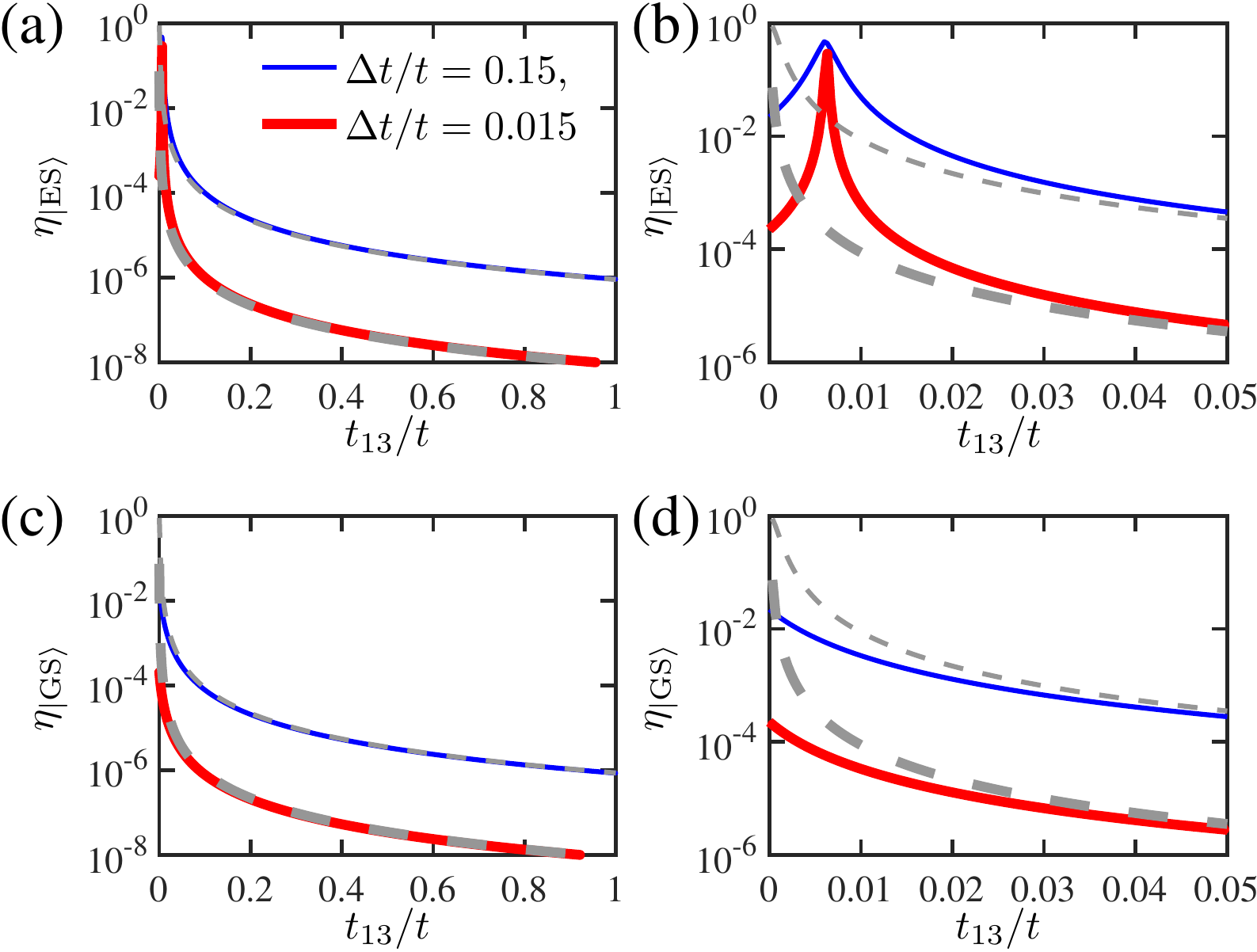} 
	\caption{Leakage as a function of tunneling rate $t_{13}$ where $|\Delta t| > 0$ for (a) $|ES\rangle$ at $\varepsilon_q=800 \mu$eV and (c) $|GS\rangle$ at $\varepsilon_q=-800 \mu $eV. The parameters are $t=10 \mu $eV, $\varepsilon_d=$ 0. Solid (dashed) line shows the numerical (analytical) solution. (b) ((d)) is a zoomed-in version of the small $\varepsilon_q$ range of (a) ((c)).}
	\label{fig:dt-nonzero}
\end{figure}

\begin{figure}[t]
	\centering
	\includegraphics[width=0.45\columnwidth]{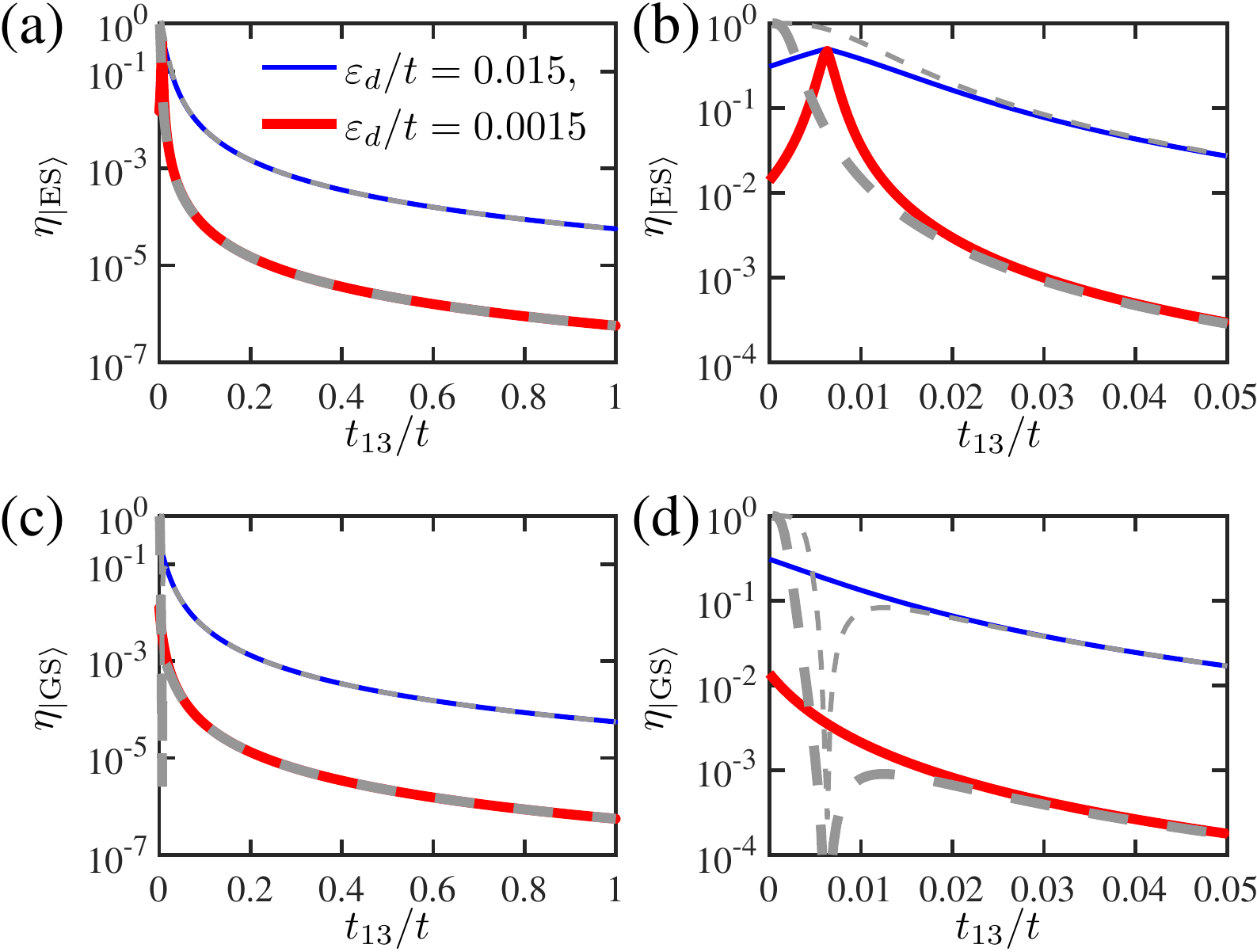} 
	\caption{Leakage as a function of tunneling rate $t_{13}$ where $|\varepsilon_d| > 0$ for (a) $|ES\rangle$ at $\varepsilon_q=800 \mu $eV and (c) $|GS\rangle$ at $\varepsilon_q=-800 \mu $eV. The parameters are $t=-10 \mu $eV and $\Delta t =$ 0. Solid (dashed) line shows the numerical (analytical) solution. (b) ((d)) is a zoomed-in version of the small $\varepsilon_q$ range of (a) ((c)).}
	\label{fig:ed-nonzero}
\end{figure}

Severe leakage happens when $|\varepsilon_q| \gg 0$. If $\varepsilon_q\rightarrow+\infty$, leakage happens between $|ES\rangle$ and $|L\rangle$, and we denote $\eta \approx\eta_{|ES\rangle} = |\langle L |ES\rangle |^2$.  If $\varepsilon_q\rightarrow-\infty$, leakage happens between $|GS\rangle$ and $|L\rangle$, $\eta \approx\eta_{|GS\rangle} = |\langle L |GS\rangle|^2$. These results are plotted, with comparison to numerical results, in Fig. \ref{fig:leakage-dt-ed-nonzero} in the main text as well as in Fig.~\ref{fig:dt-nonzero} and Fig.~\ref{fig:ed-nonzero}. Fig. \ref{fig:leakage-dt-ed-nonzero} in the main text shows the case in which both $\Delta t$ and $\varepsilon_d$ are nonzero. For completeness, we show the results of the case with $\Delta t\neq0$ and $\varepsilon_d=0$ in Supplementary Fig.~\ref{fig:dt-nonzero}, and those of the case with $\Delta t=0$ and $\varepsilon_d\neq0$ in Supplementary Fig.~\ref{fig:ed-nonzero}. Overall the behavior seen from Supplementary Figs.~\ref{fig:dt-nonzero} and \ref{fig:ed-nonzero} are very similar to that of Fig. \ref{fig:leakage-dt-ed-nonzero} in the main text, i.e. $t_{13}$ substantially reduces leakage, except that for $|ES\rangle$ the leakage first increases at very small $t_{13}$ values (because it is brought closer to $|L\rangle$ and then decreases as $|ES\rangle$ moves further away.

It is interesting to further simplify the expression of the leakage by discarding the terms of $\mathcal{O}[\alpha]^3$ or higher order (since $\alpha=\left| \frac{2t}{\varepsilon_{q}-t_{13}} \right|  \rightarrow 0$ when $|\varepsilon_{q}| \gg |t_{13}|$), the leakage probabilities can be approximated by
\begin{equation}
\begin{split}
\eta \approx \eta_{GS}^{(2)} = \eta_{ES}^{(2)} = 
\left[ \frac { \alpha \chi \left( \frac { \Delta t }{ 2t_{ 13 } }  \right)  }{ \sqrt { 4+\frac { \alpha ^{ 4 } }{ 16 } +\alpha ^{ 2 }\left( \frac { \Delta t }{ 2t_{ 13 } }  \right) ^{ 2 } }  } -\frac { \left(\frac { \varepsilon _{ d } }{ t_{ 13 } }\right)  }{ \sqrt { 4+\frac { \alpha ^{ 4 } }{ 16 } +\left( \frac { \varepsilon _{ d } }{ t_{ 13 } }  \right) ^{ 2 } }  }  \right] ^{ 2 }
\qquad \text{when }  |\varepsilon_q| \gg t_{13},
\end{split}
\label{eq:leakageapproxmation1}
\end{equation}
where the superscript $(2)$ indicates that the results is valid up to the second order in the perturbation theory. Further expanding $\alpha$ terms in the denominators and keep only up to $\mathcal{O}[\alpha]^2$, we have a simplified expression of leakage
\begin{subequations}
	\begin{align}
		\begin{split}
			\eta &=\eta_{\Delta t}+\eta_{\varepsilon_d} \approx \left( \frac { \alpha  }{ 4 }  \right) ^{ 2 }\left(\frac{\Delta t}{t_{13}}\right)^2 + \frac{1}{2}\left(\frac{\varepsilon_d}{t_{13}}\right)^2.
			\label{eq:leakageapproxmation2}
		\end{split} \\
		\begin{split}
			\frac{\eta_{\varepsilon_d}}{\eta_{\Delta t}} &\approx \frac{\frac{1}{2}\left(\frac{\varepsilon_d}{t_{13}}\right)^2}{\left( \frac { \alpha  }{ 4 }  \right) ^{ 2 }\left(\frac{\Delta t}{t_{13}}\right)^2}=\frac{8\varepsilon_d^2}{\alpha^2 \Delta t^2}
			\label{eq:leakageapproxmation3}
		\end{split} 
	\end{align}
\end{subequations}
This expression indicates that both $\Delta t$ and $\varepsilon_d$ contribute to the leakage quadratically; however the coefficient in front of $\varepsilon_d$ is much larger than $\Delta t$, indicating that for similar levels of $\varepsilon_d$
and  $\Delta t$, the former causes much more severe leakage. This also explains why in
Supplementary Figs.~\ref{fig:dt-nonzero} and \ref{fig:ed-nonzero} the leakage levels in both cases are comparable while $\Delta t$ is about 10 times $\varepsilon_d$.

\section{Details of Hund-Mulliken calculation}

The Hamiltonian of a triple-quantum-dot system (in position basis) is
\begin{equation}
\begin{split}
H  &= \sum _{ j=1 }^{ 3 }{ h(\mathbf{r}_ j) }
\end{split}
\end{equation}
where $ h(\mathbf{r}_j)$ is the single-particle Hamiltonian of particle $j$
\begin{equation}
h(\mathbf{r}_j) =
\frac{1}{2 m^*}\left( \mathbf{p}_j - e \mathbf{A}_j \right)^2+V(\mathbf{r}_j)
\end{equation}
Since there is only one electron, two-particle interactions are absent. There are three possible states (denoted as position basis $\{|100\rangle$,$|010\rangle$,$|001\rangle\}$ in the main text).

\begin{figure}[t]
	\centering
	\includegraphics[width=0.4\columnwidth]{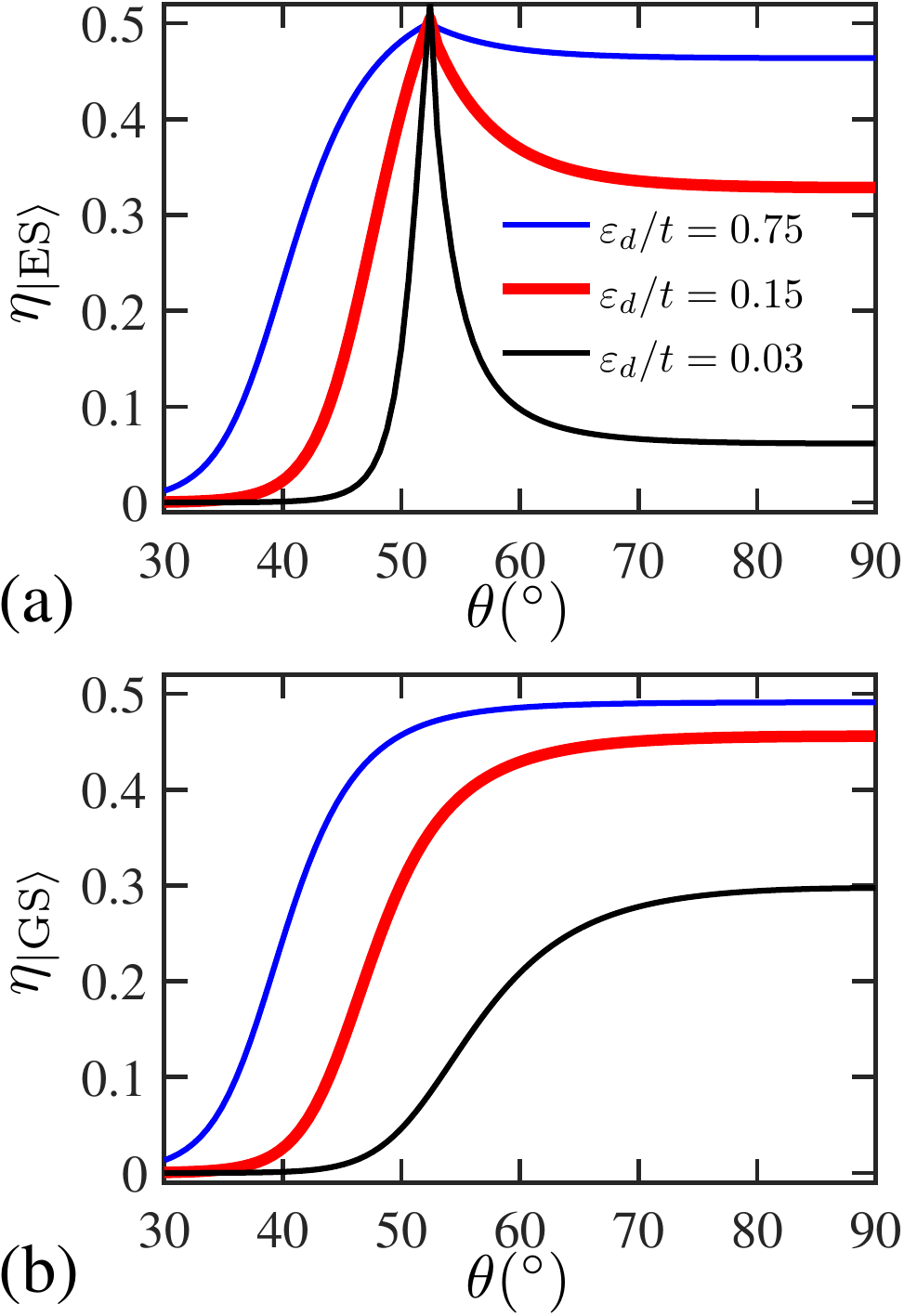}
	\caption{Leakage arises from dipole detuning noise $|\varepsilon_d|$ as a function of bending angle $\theta$ for (a) $|ES\rangle$ where $\varepsilon_q = 800 \mu $eV and (b) $|GS\rangle$ where $\varepsilon_q = -0.8 \mu $eV. The parameters are $a=150$nm, $\hbar \omega_0 = 0.7$meV, $\xi = 2.2$meV, $\xi_{13}=0$meV, $m^*=0.017m_e$ for GaAs.}
	\label{fig:leakage-HM-ed}
\end{figure}

Figure~\ref{fig:triangle-dot-figure} of the main text shows the confinement potential the electron experiences in a triple-quantum-dot device. The positions of the dots are defined as
\begin{subequations}
	\begin{align}
		\mathbf{R}_1 &= (-a \sin\theta, a \cos \theta), \\
		\mathbf{R}_2 &= (0,0), \\
		\mathbf{R}_3 &= (a \sin\theta, a \cos \theta),
	\end{align}
\end{subequations}
where $a$ is the interdot distance and $\theta$ is the bending angle. The confinement potential is defined as a sum of two parts, 
\begin{equation}
\label{confinement-potential}
V(x,y) = V_0(x,y) + G(x,y).
\end{equation}
The first part, $V_0(x,y)$, has a conventional quadratic form and is divided into three regions in x-y plane, each of which corresponds to a quantum dot,  
\begin{equation}
\begin{split}
V(x,y)  =
\begin{cases}
v_2
& \text{if } x>0 \text{ and } y< -x\tan \theta + \frac{a}{2 \cos \theta}, \\
v_3	
& \text{if } x>0 \text{ and } y> -x\tan \theta + \frac{a}{2 \cos \theta}, \\
v_2
& \text{if } x<0 \text{ and } y< x\tan \theta + \frac{a}{2 \cos \theta}, \\
v_1
& \text{if } x<0 \text{ and } y> x\tan \theta + \frac{a}{2 \cos \theta},
\end{cases}
\end{split}
\end{equation}
where 
\begin{equation}
v_j = \frac{1}{2} m^* \omega_0^2 |\mathbf{r}-\mathbf{R}_j|^2+\mu_j.
\end{equation}
The second part, $G(x,y)$,  encapsulates information of the tunable barriers $(\xi_{jk})$ between adjacent dots in triangular form of the triple dots ($\xi_{13}=0$ when the geometry of triple dots is close to lateral form),
\begin{equation}
G(x,y) = \sum _{ j<k }^{ 3 }{ G_{jk}(x,y,\mathbf{R}_{jk},\xi _{jk})}, 
\end{equation}
where  
\begin{subequations}
	\begin{align}
		G_{jk} &= \xi_{jk} \exp \left[ -\frac{32|\mathbf{r}_{jk} - \mathbf{R}_{jk}|^2}{a^2}\right],\\
		\mathbf{R}_{12} &= \left(-\frac{a}{2} \sin \theta , \frac{a}{2} \cos \theta \right),\\
		\mathbf{R}_{23} &= \left(\frac{a}{2} \sin \theta , \frac{a}{2} \cos \theta \right) \\
		\mathbf{R}_{13} &= \left(0, a \cos \theta \right)
	\end{align}
\end{subequations}

With the confinement potential defined, we can solve the Hamiltonian by adopting  the Hund-Mulliken approximation which keeps only the lowest Fock-Darwin state. The ground state is approximated to be the solution of a harmonic oscillator 
\begin{equation}
\psi_j (\mathbf{r}) = \frac{1}{a_B \sqrt{\pi}} \exp\left[ - \frac{1}{2 a_B^2} |\mathbf{r} - \mathbf{R}_j|^2\right],
\end{equation}
where $a_B = \sqrt{\hbar/(m^* \omega_0)}$ and $j=1,2,3$ refers to the three quantum dots.
The approximated single particle wave function is then obtained by orthonormalizing the three Fock-Darwin states by the following transformation
\begin{equation}
\{\psi_1,\psi_2,\psi_3 \}^\intercal = \mathcal{O}^{1/2}\{ \Psi_1,\Psi_2,\Psi_3 \}^\intercal 
\end{equation}
where $\mathcal{O}$ is the overlap matrix defined to be $\mathcal{O}_{jk} = \langle \psi_i | \psi_j \rangle$.
The matrix elements in the Hamiltonian, Eq.~\eqref{eq:ham-posBases} in the main text, can then be determined by
\begin{subequations}
	\begin{align}
		\varepsilon_j &= \langle \Psi_j | h(\mathbf{r}_j) | \Psi_j \rangle, \\
		t_{jk} &= -\langle \Psi_j | h(\mathbf{r}_j) | \Psi_k \rangle.
	\end{align}
\end{subequations}
Diagonalization of the Hamiltonian yields the eigenvalues and eigenstates for a particular set of dot parameters.

In Fig.~\ref{fig:leakage-HM-dt} of the main text we have shown the case for which $\Delta\xi$, the relative height of barriers, is varied. As noted in the main text, due to the nature of the microscopic calculation, $\Delta\xi$ not only changes $\Delta t$ but also indirectly $\varepsilon_d$. In order to complete the discussion, we show the case for which $\varepsilon_d$ is varied in the microscopic calculation (of course $\Delta t$ is at the same time indirectly changed) in Supplementary Fig.~\ref{fig:leakage-HM-ed}. The main message is similar to that from Fig.~\ref{fig:leakage-HM-dt} of the main text: For $|ES\rangle$, as $\theta$ reduces to about 50$^\circ$ the leakage first increases but then drops drastically. For $|GS\rangle$ the leakage reduces monotonically as $\theta$ is reduced from 60$^\circ$ to 30$^\circ$. Overall, having the triple quantum dots in an equilateral triangle (therefore a nonzero $t_{13}$) reduces the leakage substantially.

\section{Suppression of leakage using pulse sequence}

Leakage can be alternatively suppressed using pulse sequences as proposed in \cite{Ghosh.17}. Here we perform a comparison between the effectiveness of pulse sequences and our method, and we also study the benefit when we combine the two.

Using the language of \cite{Ghosh.17}, our Hamiltonian becomes a combination of three terms (up to an overall energy shift of $\varepsilon_{q}/2+t_{13}/2$):
\begin{equation}
H_\text{z} (\varepsilon_q) 
= 
-\left(
\begin{array}{ccc}
\frac{\varepsilon_q - t_{13} }{2} & 0 & 0 \\
0 & -\frac{\varepsilon_q - t_{13}}{2} & 0 \\
0 & 0 & 0 \\
\end{array}
\right),
\quad
H_\text{x} (t) = -\left(
\begin{array}{ccc}
0 & t & 0 \\
t & 0 & 0 \\
0 & 0 & 0 \\
\end{array}
\right),
\quad
H_\text{leak} (\varepsilon_d) = -\left(
\begin{array}{ccc}
0 & 0 & 0 \\
0 & 0 & \varepsilon_d \\
0 & \varepsilon_d & -\frac{\varepsilon_q + 3t_{13} }{2} \\
\end{array}
\right).
\end{equation}
Compared to \cite{Ghosh.17}, our Hamiltonian has an extra minus sign, but this would not affect the results.

Reference \cite{Ghosh.17} only considered noises arising from $\delta\varepsilon_d$. We denote a rotation of angle $\phi$ around axis $\hat{r}$ as $R_{\hat{r}}(\phi)$, and the rotations around $\hat{z}$ and $\hat{x}$ in a noisy  $\delta\varepsilon_d$ environment are
\begin{subequations}
	\begin{align}
		R_{\hat{\text{z}}} (\varphi) &= U_\text{z}(\varepsilon_q,\varepsilon_d,\varphi) =e^{-i\left[ H_\text{z}(\varepsilon_q) + H_{\text{leak}}(\varepsilon_d) \right]\varphi/(2\varepsilon_q)}, \\
		R_{\hat{\text{x}}} (\zeta) &= U_\text{x}(t,\varepsilon_d,\zeta) =e^{-i\left[ H_\text{x}(t) + H_{\text{leak}}(\varepsilon_d) \right]\zeta/(2t)}.
	\end{align}
\end{subequations}
An arbitrary single-qubit rotation about an axis in $xy$ plane can therefore be decomposed into elementary rotations $R_{\hat{\text{z}}} (\varphi)$ and $R_{\hat{\text{x}}} (\zeta)$. For example, we define a ``$z$-$x$-$z$'' sequence as
\begin{align}
	R_{\text{zxz}}(\varphi,\zeta) & =R_{\hat{\text{z}}}(\varphi)R_{\hat{\text{x}}}(\zeta)R_{\hat{\text{z}}}(-\varphi)\notag\\
	& = U_\text{z}(\varepsilon_q,\varepsilon_d,\varphi)U_\text{x}(t,\varepsilon_d,\zeta)U_\text{z}(-\varepsilon_q,\varepsilon_d,\varphi).
\end{align}
It has been found that when
\begin{equation}
\varepsilon_q = - \frac{t \varphi}{4} \cot\left( \frac{\zeta}{4}\right),
\label{eq:condition}
\end{equation}
The first order contribution arising from $\varepsilon_d$ vanishes in the ``bang-bang'' limit (i.e. the pulses rise/set instantaneously to the desired values) \cite{Ghosh.17}. If the pulses have a finite (nonzero) rising time, Eq.~\eqref{eq:condition} substantially reduces the first order contribution from $\varepsilon_d$ but not completely. One then needs to numerically adjust the height and duration of individual pulses in order to optimize the suppression of leakage.

As an example, we study the rotation $R_{\hat{x}+\hat{y}} (\pi/2)$ that can be decomposed as
\begin{equation}
R_{\hat{x}+\hat{y}} (\pi/2) = e^{-7i\pi/8} R_\text{xz}(-7\pi/4,\pi/2)= e^{-7i\pi/8} R_{\hat{z}}(-7\pi/4) R_{\hat{x}}(\pi/2),
\end{equation}
if no correction is desired. Nevertheless, a ``$z$-$x$-$z$'' sequence must be employed if we want to suppress the leakage:
\begin{equation}
R_{\hat{x}+\hat{y}} (\pi/2)= R_{\text{zxz}}(\pi/4,\pi/2)= R_{\hat{z}}(\pi/4) R_{\hat{x}}(\pi/2) R_{\hat{z}}(-\pi/4).
\end{equation}
We consider two cases: $t_{13} = 0$ and $t_{13} = 7.5\ \mathrm{GHz}\approx 30\ \mu$eV with $t$ fixed at 3 GHz. We also take into account the finite rise time of $50$ ps for every pulse \cite{Ghosh.17}. Numerical optimization is performed with respect to the fidelity defined as
\begin{equation}
F = \frac{1}{12}\left[\text{Tr}\left( \mathcal{U} \mathcal{U}^\dagger \right) + \left|\text{Tr}\left( U^\dagger_{\text{target}} \mathcal{U} \right)\right|^2\right],
\label{eq:fidelity}
\end{equation}
where $U_{\text{target}}$ is the desired operation, and $\mathcal{U}$ the actual evolution.
Leakage probabilities $\eta$ are defined as $\eta_{|GS\rangle}=|\langle L | \mathcal{U} (t;t=0)|GS \rangle _{t=0}|^2$ where $|GS\rangle_{t=0} = |E\rangle$, and $\eta_{|ES\rangle}=|\langle L | \mathcal{U} (t;t=0)|ES \rangle _{t=0}|^2$ where $|ES\rangle_{t=0} = |C\rangle$.

\begin{figure} [t]
	\centering{
		\includegraphics[width=0.65\columnwidth]{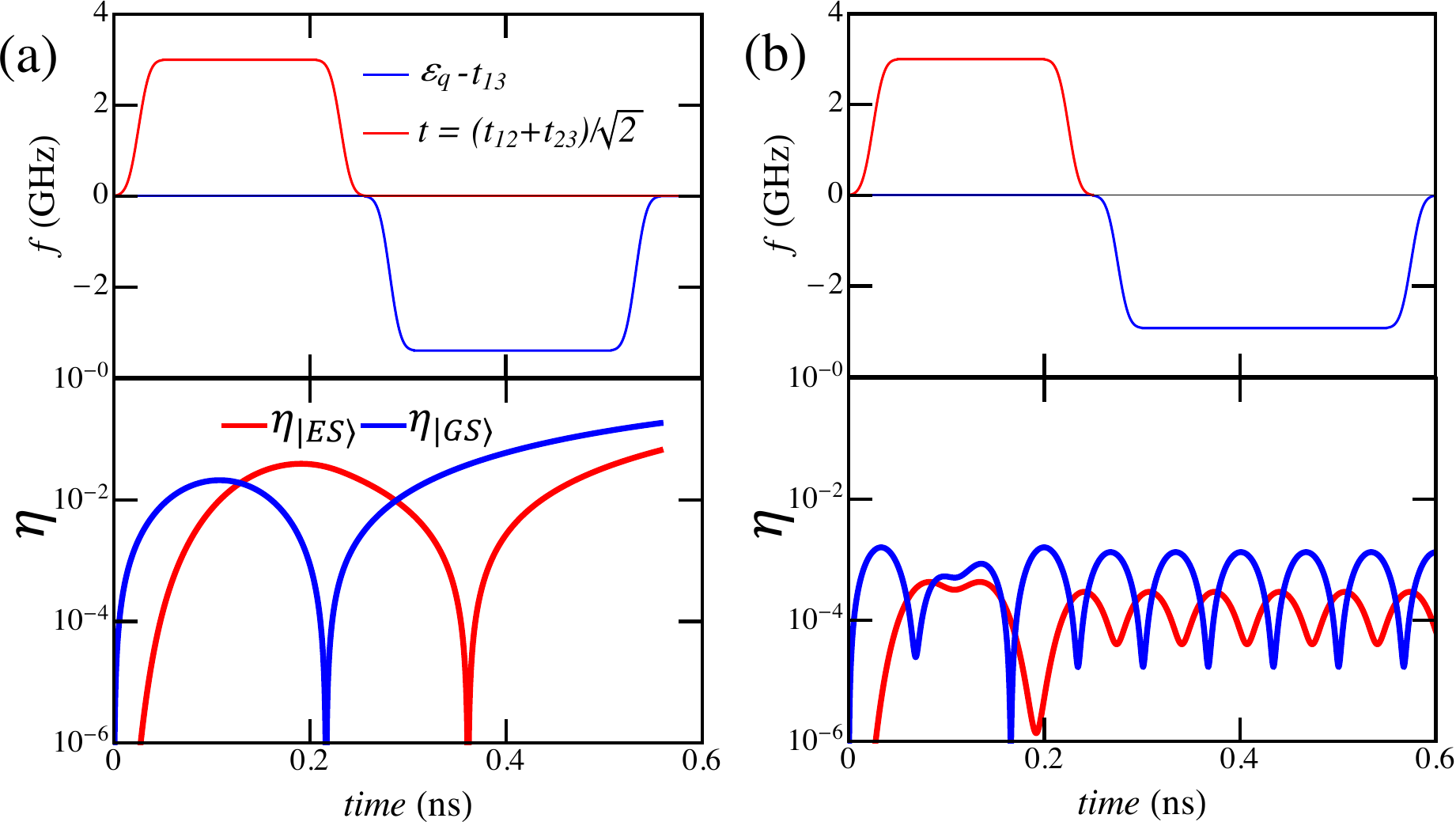}
		\includegraphics[width=0.65\columnwidth]{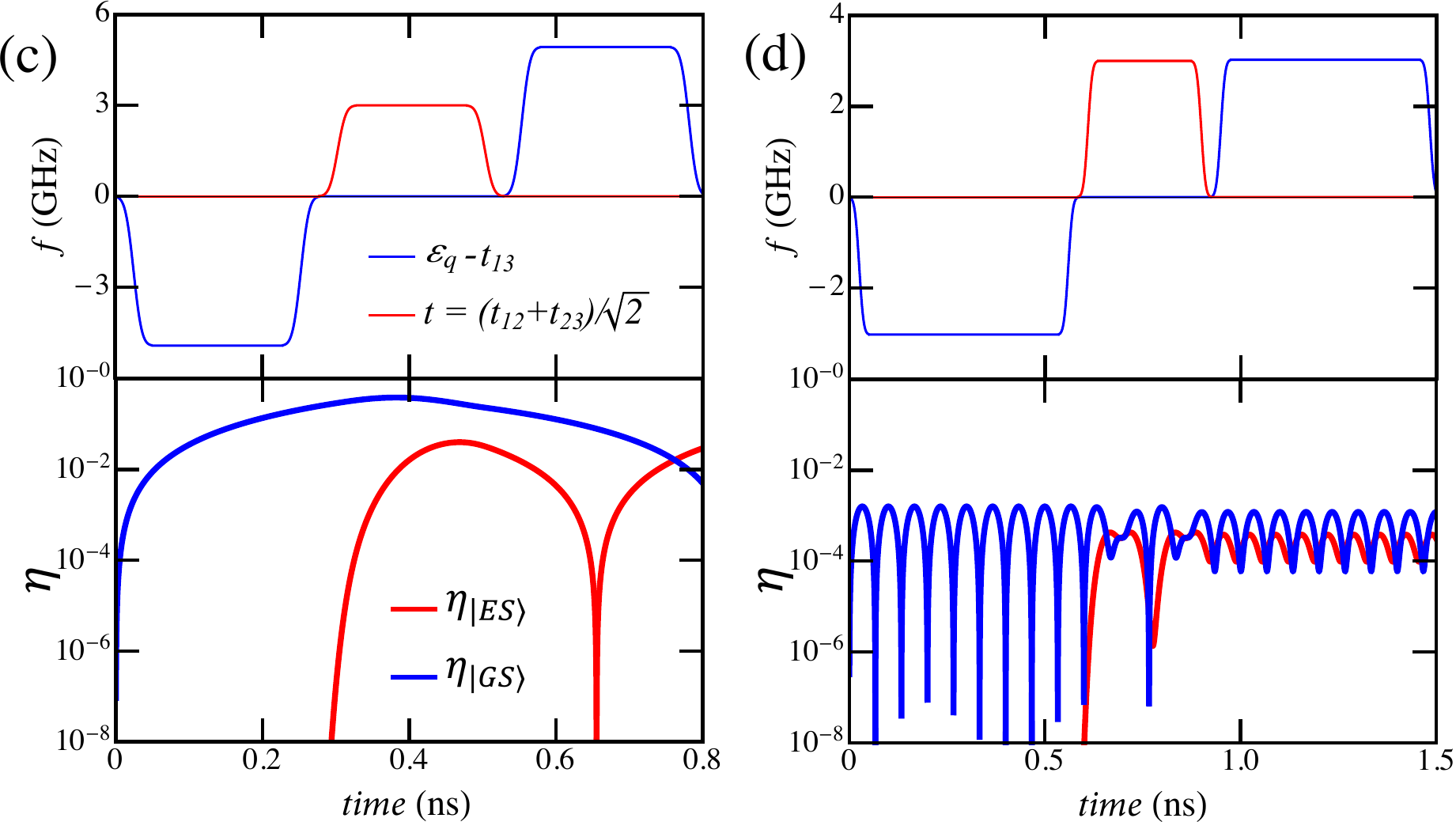}
		\caption{Numerically optimized gate operation for $R_{\hat{x}+\hat{y}} (\pi/2)$ at $\varepsilon_d = 0.3$ GHz. (a), (b): The pulse sequence and leakage probabilities for naive two-pulse sequence, $R_{\text{xz}}(\varphi,\zeta) = R_{\hat{\text{x}}}(\varphi)R_{\hat{\text{z}}}(\zeta)$. (c), (d): The pulse sequence and leakage probabilities for $z$-$x$-$z$ sequence (with noise correction), $R_{\text{zxz}}(\varphi,\zeta) = R_{\hat{\text{z}}}(\varphi)R_{\hat{\text{x}}}(\zeta)R_{\hat{\text{z}}}(\varphi)$. (a), (c): $t_{13} = 0$. (b), (d): $t_{13} =7.5$ GHz. In each panel, the top subfigure shows the pulse sequence with $f$ denoting the pulse amplitude, and the bottom subfigure shows the leakage probabilities.}
		\label{fig:pulseSeqLeakage}
	}
\end{figure}

Numerical results are shown in Fig.~\ref{fig:pulseSeqLeakage}. It can be observed that for the naive pulse sequences ($R_{\text{xz}}$), the leakage simply accumulates throughout the gate operation, ending up with a large leakage at the conclusion of the sequence. [Fig.~\ref{fig:pulseSeqLeakage}(a)]. When the composite pulses are employed, the leakage first accumulates but then decreases, and the leakage for the gate is smaller compared to the previous case [Fig.~\ref{fig:pulseSeqLeakage}(c)]. In both cases, applying a nonzero $t_{13}$ reduces the leakage substantially. For naive sequences the results shown in Fig.~\ref{fig:pulseSeqLeakage}(b) have a leakage level two orders of magnitude lower than that of Fig.~\ref{fig:pulseSeqLeakage}(a). For corrected composite sequences, the nonzero $t_{13}$ reduces the leakage by about one order of magnitude [compare Fig.~\ref{fig:pulseSeqLeakage}(d) and Fig.~\ref{fig:pulseSeqLeakage}(c)].


\begin{thebibliography}{27}
\bibitem [{lea()}]{leak}%
	\BibitemOpen
	\href@noop {} {}\bibinfo {note} {Since this disagreement happens for a very
		small $t_{13}$ and as $t_{13}$ is further increased, the leakage is
		substantially reduced as desired, we will not perform a calculation with
		degenerate perturbation theory to resolve the disagreement, as that is
		unimportant for our main results.}\BibitemShut {Stop}%
\end{thebibliography}
\end{document}